\documentclass{PoS}

\usepackage{gensymb}
\usepackage{graphicx}
\usepackage{amsmath}
\usepackage{amssymb}
\usepackage{tabularx}
\usepackage{rotating}
\usepackage{float}
\usepackage[caption=false]{subfig}
\usepackage{multicol}

\title{Recent Results of the VERITAS Galactic Science Program}

\ShortTitle{VERITAS Galactic Highlights}

\author{\speaker{Gregory T. Richards}\\
        Department of Physics and Astronomy and the Bartol Research Institute, University of Delaware, Newark, DE 19713.\\
        E-mail: \email{grichard@udel.edu}}

\author{The VERITAS Collaboration\footnote{for collaboration list see PoS(ICRC2019)1177}\\
        \email{veritas.sao.arizona.edu}}

\abstract{VERITAS (Very Energetic Radiation Imaging Telescope Array System) is one of the most sensitive currently operating arrays of imaging atmospheric Cherenkov telescopes, which detect very high-energy (VHE; E > 100 GeV) gamma rays. VERITAS is currently in its 11th year of full-array operations with four 12m-diameter telescopes. Many Galactic sources of VHE gamma rays have been detected by VERITAS, such as pulsar wind nebulae, binary systems, and supernova remnants, and the study of VHE emission from these objects has enabled a deeper understanding of the underlying physical processes responsible for the observed gamma rays. Recent highlights from the VERITAS Galactic science program will be presented, including results on pulsar searches, follow-up of sources detected by HAWC, and the 50-year-period binary PSR J2032+4127.}

\FullConference{36th International Cosmic Ray Conference -ICRC2019-\\
		July 24th - August 1st, 2019\\
		Madison, WI, U.S.A.}

\begin{document}

\section{Introduction}
The Galactic VHE sky contains objects of a variety of classes, including pulsar wind nebulae (PWNe), binary systems, supernova remnants (SNRs), and pulsars.  Many Galactic gamma-ray sources are resolved as extended in the VHE band, which enables morphological studies and derivation of spatially-dependent spectra.  Studying the characteristics of the emission from these objects allows to determine the locations and mechanisms of the particle acceleration responsible for the observed radiation.   This proceeding provides an overview of recent Galactic science results of the VERITAS collaboration.

\section{VERITAS}
VERITAS is an array of four 12\,m diameter IACTs and is located at the Fred Lawrence Whipple Observatory in southern Arizona (31$^{\circ}$ 40' N, 110$^{\circ}$ 57' W, 1.3\,km a.s.l.). VERITAS started full-array 
operations in 2007.  The telescope reflectors consist of 345 hexagonal mirror facets, and the cameras comprise 499 photomultiplier tubes giving a field of view (FoV) of $\sim$3.5$^{\circ}$.  VERITAS is sensitive to VHE gamma-ray photons in the energy range $0.85$ to $>$30\,TeV with a sensitivity to detect a 1\% Crab Nebula source in $\sim$25 hr.  It has an angular resolution of $0.1^{\circ}$ at 68\% containment and a pointing-accuracy error of less than 50 arcseconds~\cite{2008AIPC.1085..657H}.  The VERITAS data analysis results presented here follow the methodology outlined in~\cite{2008ApJ...679.1427A}.

%VERITAS is an array of four, $12\U{m}$ diameter atmospheric Cherenkov telescopes located at the Fred Lawrence Whipple Observatory in Arizona \cite{Holder06,Park15}. The array is approximately square, with side lengths of $\sim100\U{m}$. Each steerable telescope is equipped with a 499-pixel photomultiplier tube camera, covering a circular field-of-view with a diameter of $3.5\degree$. For observations in the gamma-ray band, images of the Cherenkov light from cosmic ray and gamma-ray initiated air showers are recorded and processed using standard tools \cite{Maier17,Daniel_VEGAS}. Digitization of the AC-coupled photomultiplier tube signals occurs at 500 megasamples-per-second, for Cherenkov images which exceed the array trigger thresholds. The readout is restricted to a 20\U{ns} window around the time of the trigger. 
%The requirements for optical photometry are significantly different to gamma-ray observations, and so a specialized data acquisition chain is used which is fully parallel to normal Cherenkov operations. Details are given elsewhere in these proceedings \cite{Daniel_ECM} but, briefly, the DC-coupled photomultiplier tube voltages are recorded using a commercial 14-bit data logger (DATAQ instruments DI-710-ELS) on up to 16 channels for each telescope. The system can record on two channels at 2,400\U{Hz}, or on 16 channels at 300Hz.

\section{The Binary System PSR J2032+4127/MT91 213} \label{sec:j2032}

%What is the system

PSR J2032+4127/MT91 213 is a binary system comprising the young gamma-ray pulsar J2032+4127 and a 15\,M$_{\odot}$ Be star with an orbital period of $\sim$50\,yr~\cite{2015MNRAS.451..581L}.  The binary nature of the system PSR J2032+4127/MT91 213 (hereafter referred to as J2032) was only recently established via radio observations of dramatic changes to the pulsar spin-down rate~\cite{2015MNRAS.451..581L}.  The system experienced periastron passage in 2017 November, where the orbital separation fell to just  $\sim$0.5\,AU~\cite{2017ApJ...836..241T}.    %indicating Doppler shifting due to binary motion.
Co-located with J2032 is the extended VHE gamma-ray source TeV J2032+4130, which was long-classified as an unidentified object, despite thirteen years of observations since its discovery.  It is now thought likely to be a pulsar wind nebula, given that it is co-located with the young, energetic \textit{Fermi}-LAT-detected pulsar J2032+4127.

The X-ray flux from J2032 was seen to rise for some time, with an increase by a factor of 70 between 2002 and 2016 reported in~\cite{2017MNRAS.464.1211H}.  This brightening has been interpreted as a result of increasing wind energy in the shock region formed by the pulsar and stellar winds~\cite{2017MNRAS.464.1211H}.  More recently, \cite{2018ApJ...857..123L} reported a continuing X-ray flux increase up to 2017 May in \textit{Swift}-XRT data.  Evidence for variability in the X-ray light curve is also present.   In contrast to the increasing X-ray flux from J2032, the high-energy gamma ray flux seen in \textit{Fermi}-LAT data appeared steady leading up to and during the 2017 periastron passage, likely due to masking by the strong magnetospheric emission~\cite{2018ApJ...857..123L}.  %Further, the light curve is consistent with a uniform flux; i.e., no variability is detected.

\begin{figure}[t]
  \centering
  \subfloat[]{\includegraphics[scale=0.355]{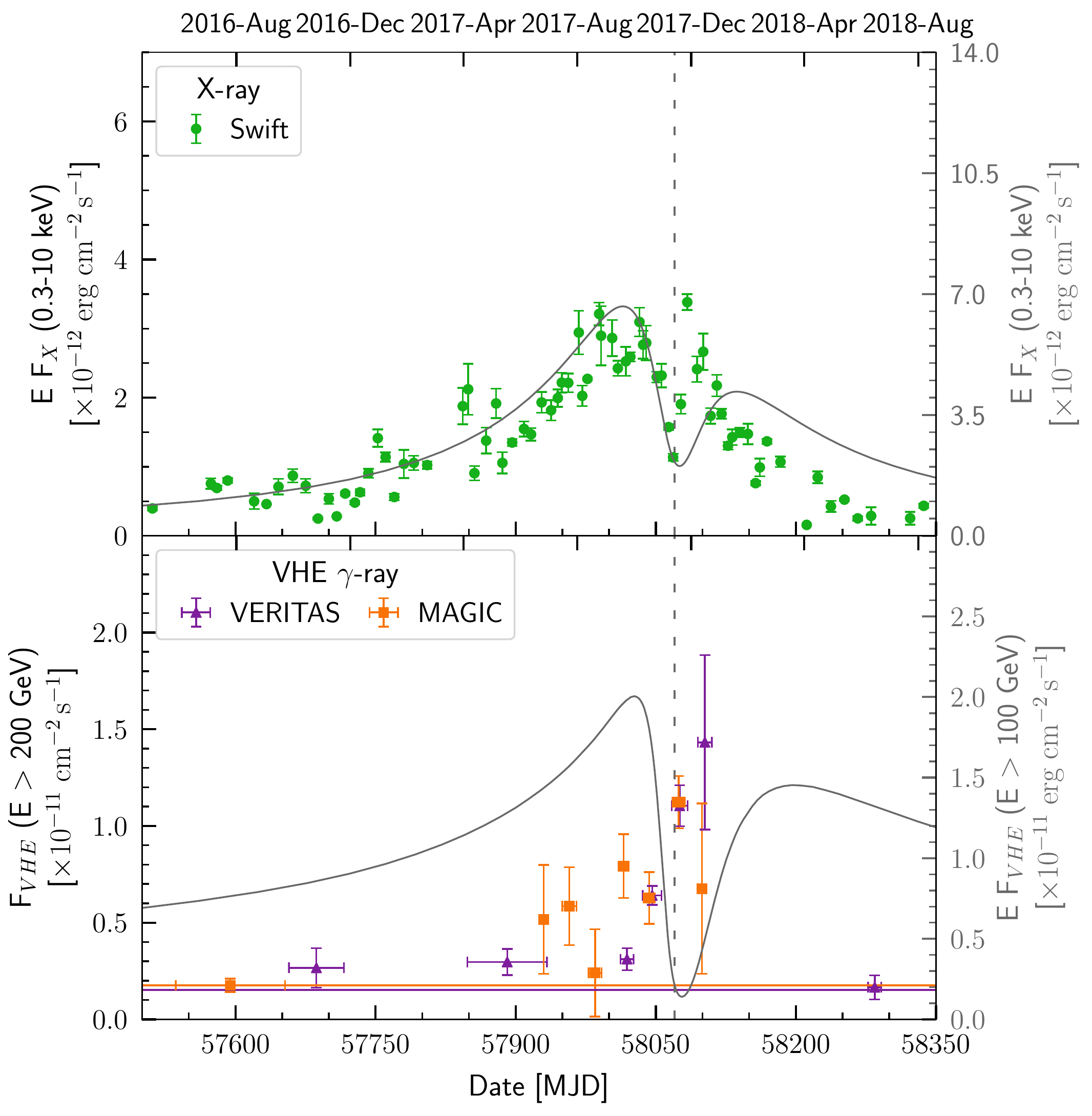}}   
  \subfloat[]{\includegraphics[scale=0.355]{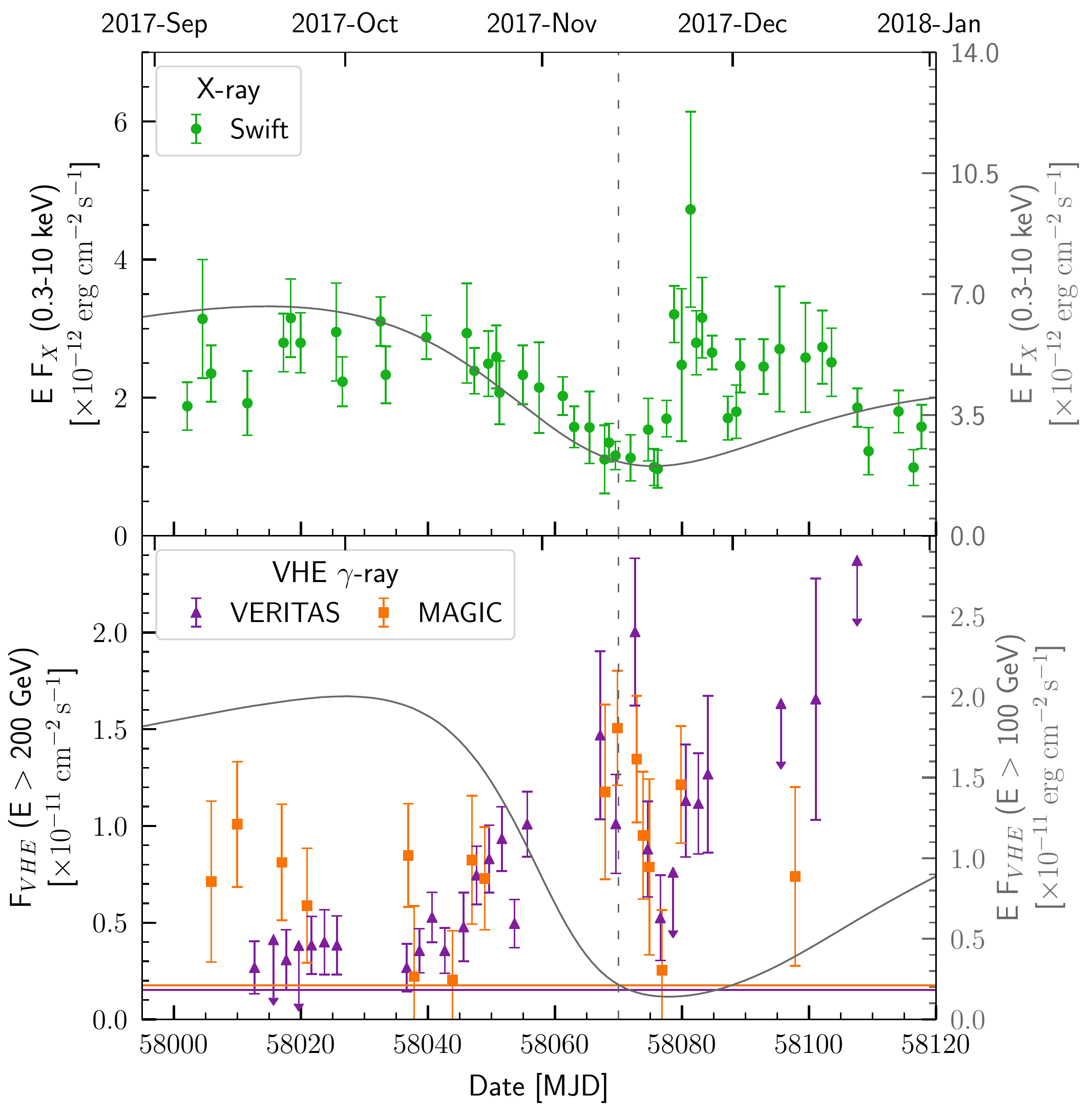}} \\
 \caption{Light curves of PSR J2032+4127/MT91 213 for the full data set (left) and near periastron (right).  Upper panels show the 0.3--10.0 keV {\it Swift-XRT} light curves of PSR J2032+4127/MT91 213.  Lower panels show the $>200$\,GeV light curves from VERITAS (green) and MAGIC (blue). The average fluxes seen by VERITAS and MAGIC prior to 2017 are indicated by horizontal solid lines.  The solid gray lines (right axes) are the energy-flux light curve predictions from~\cite{} for X-rays and updated predictions from~\cite{2018ApJ...857..123L} using parameters from~\cite{2017ApJ...836..241T} (J. Takata 2018, priv. comm.) for VHE gamma-rays, assuming an inclination angle of $60^{\circ}$. The vertical gray dashed line indicates the time of periastron passage.}
 \label{fig:2032_lightcurves}
\end{figure}

Recent observations by VERITAS and MAGIC revealed a rising VHE gamma-ray flux beginning in 2017 September associated with the emergence of a new point source in the TeV J2032+4127 region~\cite{2018ApJ...867L..19A}.  The emergence of this new point source confirmed that the system is a gamma-ray binary and resulted in confirmation of the second-known gamma-ray binary with a known compact companion\footnote{The other is PSR B1259--63/LS 2883.}.  Long-term and near-periastron VHE and X-ray light curves of J2032 are shown in Figure~\ref{fig:2032_lightcurves}.  The VHE flux displays a clear rise and then a dip as the system nears and passes periastron, and the VHE flux is poorly correlated during this time with the overlaid model prediction.  That the time-dependent behavior of the VHE flux is not well modeled at present underscores the poorly understood geometry of the system---significant revisions on this front will be required moving forward~\cite{2018ApJ...867L..19A}.        %That TeV J2032+4130 is an extended galactic object at VHEs would typically implicate the presence of a PWN as suggested in~\cite{2014ApJ...783...16A}, though this interpretation was brought into question due to the considerably hard VHE spectrum and the recently identified binary.

%Updated VHE gamma-ray results were recently presented by VERITAS in~\cite{2017arXiv170804718B} and by VERITAS and MAGIC in a joint ATel (\#10810) published 3 October 2017~\cite{2017ATel10810....1}.  The VERITAS data in~\cite{2017arXiv170804718B} comprised a total of 45\,hr with 17.6\,hr taken in fall 2012, 8.5\,hr in fall 2016\, and 11.0\,hr in spring 2017.  A test for an elevated flux at the pulsar location in yearly bins did not reveal any concrete evidence for an increasing flux.  However, in observations conducted by VERITAS and MAGIC (15.3\,hr and 5.6\,hr, respectively) in fall 2017, both instruments observed a flux elevated by a factor of $\sim$2 compared to spring/summer 2017~\cite{2017ATel10810....1}, with both VERITAS and MAGIC reporting a clear detection of a point source at the pulsar location. The source is added to the VERITAS / MAGIC catalogs as VER J2032+414 / MAGIC J2032+4127, respectively.  The emergence of a point source as J2032 nears periastron confirms that the system is a gamma-ray binary, and this detection makes it just the second gamma-ray binary with a known compact companion\footnote{The other is PSR B1259--63/LS 2883.}.

%Our past results from ICRC

%The VERITAS results for J2032 presented herein are an update to those presented in~\cite{2017arXiv170804718B}, with the addition of new data taken in fall 2017 that was used in forming the results presented in ATel \#10810.  The analysis methodology is the same as in~\cite{2017arXiv170804718B}. 

%In 2017 September, VERITAS accumulated a total of 15.3\,hr of new data on the J2032 location.  Analysis of the September data set yields a detection significance of 6.8$\sigma$ at the location of PSR J2032+4127.  The September VHE gamma-ray flux $>$200\,GeV is measured to be $(4.2 \pm 0.7) \times 10^{-12}$\,cm$^{-2}$\,s$^{-1}$.  Analysis of $\sim$21\,hr of data taken in spring 2017 gives a flux of $(2.1 \pm 0.7) \times 10^{-12}$\,cm$^{-2}$\,s$^{-1}$ above the same energy threshold, hence the observed flux increase is a factor of $\sim$2.  Significance maps derived from the spring and fall data sets are shown in Figure~\ref{fig:2032_sigmaps}.  A clear detection of a point source at the pulsar location can be seen in the second panel.  %Since J2032 lies within the extended VHE source TeV J2032+4130, some contamination from the extended source is expected.  

%VERITAS will continue to monitor J2032 through periastron passage, and a detailed study is the subject of an upcoming VERITAS collaboration publication.
%%%%%%%%%%%%%%%%%%%%%%%%%%%%%%%%%%%%%%%%%%%%%%%%%%%%%%%%%%%%%%%%%%%%%%%

\section{The Supernova Remnant Cassiopeia A} \label{sec:casA}

%One of the primary historical motivations for VHE gamma-ray astronomy has been in understanding the origin of the "knee" feature in the cosmic-ray spectrum around $10^{15}$\,eV (1\,PeV).  
Observations of cosmic rays have long shown a contribution from within the Galaxy up to energies of $>$10$^{15}$\,eV (1\,PeV) where the spectrum shows a break at the ``knee'', above which the cosmic-ray contribution is thought to be primarily extragalactic.  Determining the Galactic source population (the so-called ``PeVatrons'') responsible for these PeV-scale cosmic rays has been one of the primary historical motivations for VHE gamma-ray astronomy, though these efforts have been fruitless with perhaps the exception of the Galactic Center~\cite{2016Natur.531..476H}.  %As discussed previously in Section~\ref{sec:gc}, the so-called Galactic "PeVatrons" responsible for cosmic-ray acceleration up to the knee have to date eluded a firm observational confirmation, %due to lack of evidence of hadronic acceleration to the required energies, with the exception of the Galactic Center~\cite{2016Natur.531..476H} (see Section~\ref{sec:gc}).  
One of the most intensively studied classes of Galactic VHE gamma-ray sources are the supernova remnants (SNRs), which have been historically proposed as potential PeV-scale accelerators of hadrons.  Diffusive shock acceleration remains the accepted explanation for cosmic-ray acceleration in SNRs and can in principle provide the required PeV-scale cosmic-ray energies~\cite{2001RPPh...64..429M}.  However, the observed VHE gamma-ray spectra of SNRs display cutoffs at sub-PeV energies~\cite{2013APh....43...71A} and thus do not currently allow the conclusion that the known population of VHE-emitting SNRs constitutes the missing PeVatrons. 

\begin{figure}[h]
  \centering
  \includegraphics[scale=0.85]{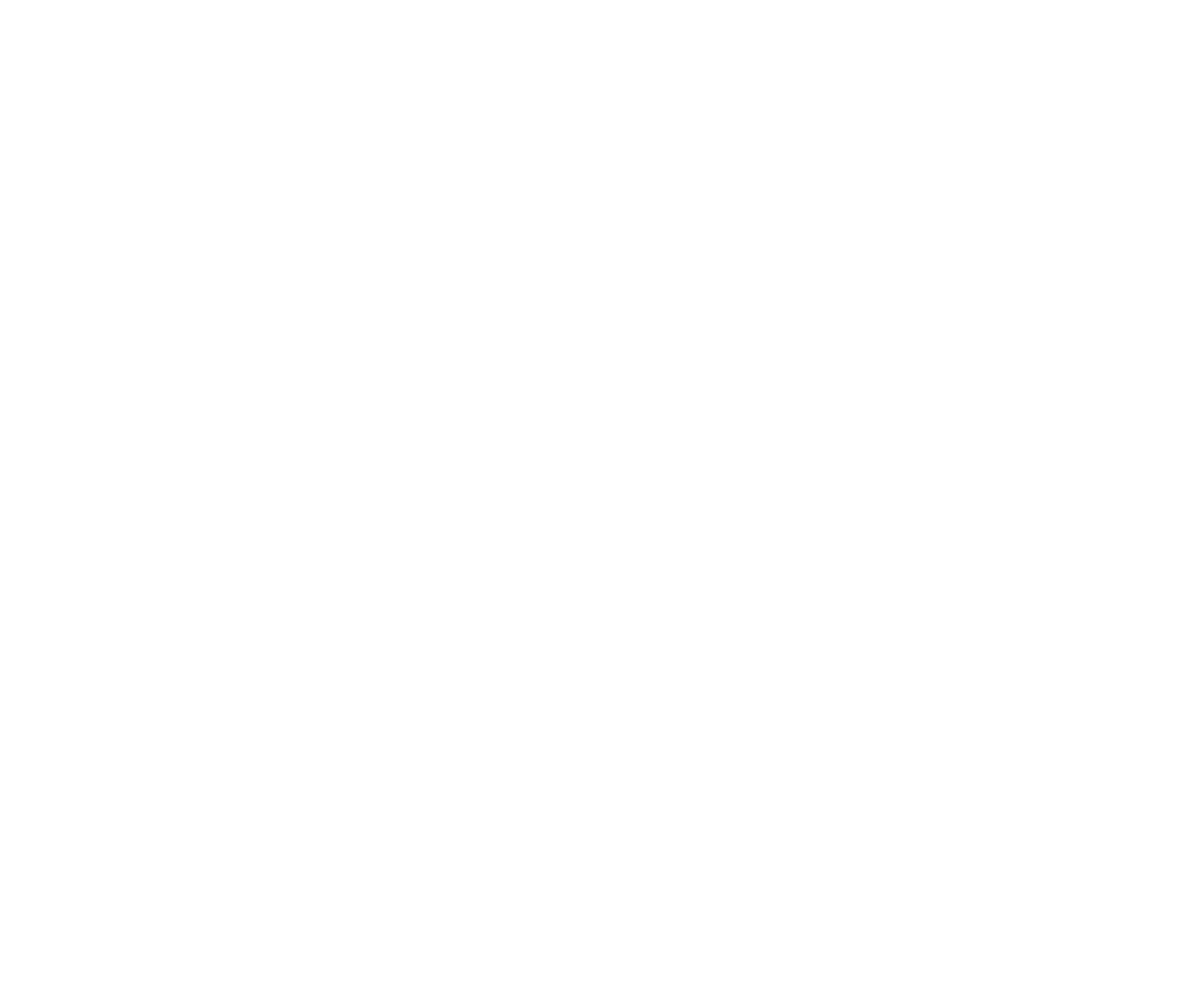}
 \caption{Combined {\it Fermi}-LAT and VERITAS SED of Cas A. The measured \textit{Fermi}-LAT spectral points are shown in red, while the VERITAS points are shown in blue.  The best-fit model (a power law with exponential cutoff) is shown with a dotted blue line. The blue shaded region represents the $1\sigma$ statistical error band on the fitted spectral model.}
 \label{fig:casAspectrum}
\end{figure}

The supernova remnant Cassiopeia A (henceforth Cas A) is one of the best-studied young SNRs in the Galaxy.  Non-thermal X-ray emission has been observed that indicates a population of multi-TeV electrons in the forward shock~\cite{2001ApJ...552L..39G}, while the observed gamma-ray emission from MeV--TeV energies is of more ambiguous origin~\cite{2014A&A...563A..88S}.  Gamma-ray observations in the VHE band by MAGIC revealed a spectrum that is best described by a power law with an exponential cutoff at $3.5$\,TeV, indicating that Cas A is not currently operating as a PeVatron.  A combined \textit{Fermi}-LAT and VERITAS spectrum also strongly favors a cutoff at a few TeV (see Figure~\ref{fig:casAspectrum}), and recent modeling tends to favor hadronic scenarios for the observed gamma-ray spectra (Abeysekara et al., {\it in prep}).  In short, though a hadronic origin of the cosmic rays in Cas A is currently preferred, VHE observations by MAGIC and VERITAS have not established Cas A as a potential source of cosmic rays up to the knee. % something about an upcoming VERITAS publication?      

\section{Follow-up of VHE Sources Discovered by HAWC} \label{sec:hawc}

%Synergy between fermi and veritas and hawc (adjacent overlapping e ranges, survey vs pointed)

The High-Altitude Water Cherenkov (HAWC) telescope provides a wide-field survey of the northern sky at TeV energies.  The second HAWC source catalog (2HWC)~\cite{2017ApJ...843...40A} contains 16 newly detected VHE gamma-ray sources that are over $1^{\circ}$ away from known sources.  Of these 16, the locations of 11 appear in archival VERITAS data collected between 2007 and 2015, while VERITAS has additionally observed the locations of other 2HWC sources with dedicated observations, bringing the total observed locations to 13.  The VERITAS analysis resulted in the detection of one of the new VHE sources: 2HWC J1953+294 (VER J1952+294)~\cite{2018ApJ...866...24A}.  A \textit{Fermi}-LAT analysis was also conducted in the energy range 10--900\,GeV, which resulted upper limits for both a point-source and extended-source search~\cite{2018ApJ...866...24A}.   

\begin{figure}[t]
  \centering
  \subfloat{\includegraphics[scale=1.00]{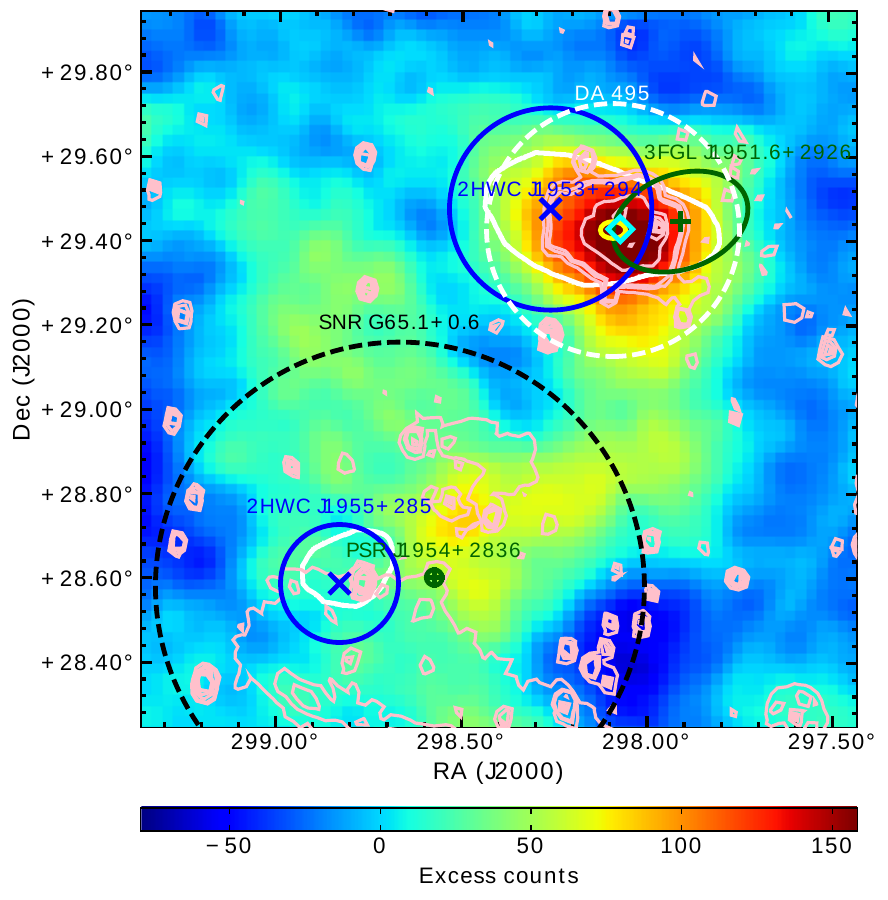}}   
  %\subfloat[]{\includegraphics[scale=0.85]{./da495_spectra_preliminary}} \\
 \caption{VERITAS VHE gamma-ray map of the DA 495 region.  The blue circles indicate the 1$\sigma$ locations of the two 2HWC sources in this region, with the blue x indicating the centroids.  Sources from the \textit{Fermi}-LAT third source catalog are shown in green.  Radio contours for DA 495 from~\cite{2003AJ....125.3145T} are drawn in pink, while the solid white curves indicate the 5$\sigma$ HAWC source locations.  The dashed white circle shows the size of the angular cut used in the VERITAS extended source search, and the dashed black circle indicates the extent of the radio emission of SNR G65.1+0.6.  More details are provided in~\cite{2018ApJ...866...24A}.}
 \label{fig:da495}
\end{figure}

One region observed by VERITAS contains two of the new VHE sources in the 2HWC: 2HWC J1953+294 and 2HWC J1955+285~\cite{2018ApJ...866...24A}.  VERITAS has accumulated a total of 64\,hr of observations of this region, which resulted in a 5.2$\sigma$ detection of a source (VER J1952+294) coincident with 2HWC J1953+294 and a non-detection of emission from the 2HWC J1955+285 location.  The likely association of 2HWC J1953+294 and VER J1952+294 is the radio pulsar wind nebula (PWN) DA 495~\cite{2018ApJ...866...24A}.  The VERITAS counts map for the DA 495 region is shown in Figure~\ref{fig:da495}. 

%The multiwavelength gamma-ray spectrum for the DA 495 region is shown in Figure~\ref{fig:da495}.  Though no firm statement can be made about the HAWC spectrum alone for this PWN, the combined \textit{Fermi}-VERITAS-HAWC spectrum favors a ...~\cite{}

The possibility of this type of multiwavelength study in gamma rays across nearly seven decades in energy underscores the value of combined \textit{Fermi}-LAT, VERITAS, and HAWC studies. While the \textit{Fermi}-LAT and HAWC continually survey the sky, 
%which naturally leads to new gamma-ray source detections.  
the high angular resolution of an IACT such as VERITAS allows detailed morphological studies in follow-up observations.  Furthermore, spectra derived from data collected by these three instruments can be combined to produce a more complete picture of the gamma-ray radiation from a variety of sources.   

%For further details about VERITAS and \textit{Fermi}-LAT observations of the new VHE sources detected by HAWC, please see~\cite{2017arXiv170805744P} and the upcoming \textit{Fermi}-VERITAS-HAWC joint publication (in preparation).

\section{Searches for VHE Gamma-Ray Pulsars } \label{sec:pulsars}

% edit all this too / chop it down!
Since the unexpected detection of the Crab pulsar in VHE gamma rays by VERITAS~\cite{2011Sci...334...69V} and MAGIC~\cite{2008Sci...322.1221A, 2016A&A...585A.133A}, one question of great interest in VHE astrophysics has been whether or not the Crab pulsar is the sole VHE-emitting pulsar.  The VHE spectrum of the Crab pulsar has been measured to be consistent with a pure power law up to 1.5\,TeV~\cite{2016A&A...585A.133A} by MAGIC, which has allowed stringent constraints to be made on the mechanism and location of the particle acceleration responsible for the emission.  In the time since the detection of the Crab pulsar, the Vela pulsar was recently detected up to $\sim$100\,GeV by H.E.S.S. II~\cite{2018A&A...620A..66H} and at energies above a few TeV by H.E.S.S.~\cite{arache_vela}, which indicated that pulsed VHE emission may be a more ubiquitous feature of energetic pulsars and not unique to the Crab.  Detecting more pulsars in the VHE band in general will of course help to further elucidate the nature of pulsed VHE emission, therefore pulsar observations continue to be of interest in gamma-ray science. 

As members of the young gamma-ray pulsar population, both the Crab and Vela are very highly ranked according to a so-called ``observability metric'' $\dot E/d^2$,\footnote{This metric give the total power output of a pulsar, $\dot E$, weighted by the inverse square of its distance.} taking the number one and two spots of all known gamma-ray pulsars.  Given that the Crab and Vela are the only pulsars known to emit at TeV energies, one natural starting point for a search for more VHE pulsars is to sort observable pulsars according to $\dot E/d^2$.

VERITAS has incidentally observed the locations of many of the top northern-hemisphere pulsars according to $\dot E/d^2$ rank~\cite{2019ApJ...876...95A}.  Such pulsars that VERITAS has observed (primarily while targeting a PWN or supernova remnant) are listed in Table~\ref{tab:pulsar_properties}, along with some properties and the VERITAS exposure time for each.  There are 13 total pulsars, and this list contains eight of the top twelve pulsars located in the northern sky when ranked in $\dot E/d^2$.  Two of the top twelve are the Crab and Geminga pulsars, which have already been observed by VERITAS~\cite{2011Sci...334...69V, 2015ApJ...800...61A}.  Although none of the 13 pulsars probed for pulsed emission in this study were detected, the derived upper limits constrain a flux that is in many cases below the flux level of the Crab pulsar, so the general statement can be made that VHE pulsed emission from each pulsar, if present, must be more faint than that observed from the Crab pulsar ($\sim$1\% Crab Nebula level)~\cite{2019ApJ...876...95A}.

\begin{table}
\centering

\caption{Properties of the 13 pulsars searched for pulsed emission by VERITAS as reported in~\cite{2019ApJ...876...95A}.  Columns 2 and 3 give the pulsar period, $P$, and time derivative of the period, $\dot P$.  The spin-down luminosities ($\dot E$) are given in column 4.   Column 5 lists the ranking in $\dot E / d^2$ for the Northern Hemisphere pulsar population.  Column 6 gives the possible PWN counterparts of the pulsars, and columns 7 and 8 give the VERITAS exposure times and average zenith angles of observations.  Values for $P$, $\dot P$, and $\dot E$ have been taken from the second {\it Fermi}-LAT pulsar catalog~\cite{2013ApJS..208...17A} unless otherwise noted. The possible PWN counterparts are taken from SIMBAD or TeVCat.
}
\vspace{3mm}
\scriptsize
\begin{tabularx}{1.0\textwidth}{cccccccc}

  \hline
   Pulsar & $P$ (ms) & $\dot P$ ($10^{-15}$) & $\dot E$ ($10^{34}$\,erg\,s$^{-1}$) & $\dot E / d^2$ Rank & Counterpart? & VTS Exposure (hr) & $\bar \theta_{\textrm{zenith}}$ ($^{\circ}$) \\ \hline

J0007+7303 &  315.9 & 357 & 44.8 & 9 & CTA 1 & 32.4 & 42 \\
J0205+6449 &  65.7 & 190 & 2644 & 3 & 3C 58 & 22.2 & 35 \\
J0248+6021 &  217.1 & 55.0 & 21.2 & 12 & - & 45.9 & 32 \\
J0357+3205 &  444.1 & 13.1 & 0.6 & 14 & - & 7.92 & 14  \\
J0631+1036 &  287.8 & 104 & 17.3 & 10 & - & 2.79 & 26 \\
J0633+0632 &  297.4 & 79.6 & 11.9 & - & - & 108 & 29 \\
J1907+0602 &  106.6 & 86.7 & 282 & 8 & MGRO J1908+06 & 39.1 & 28 \\
J1954+2836 &  92.7 & 21.2 & 105 & - & - & 5.18 & 16 \\
J1958+2846 &  290.4 & 212 & 34.2 & - & - & 13.9 & 10 \\
J2021+3651 &  103.7 & 95.6 & 338 & 4 & Dragonfly Nebula & 58.2 & 18 \\
J2021+4026 &  265.3 & 54.2 & 11.4 & 13 & $\gamma$ Cygni & 20.6 & 21 \\
J2032+4127 &  143.2 & 20.4 & 15~\cite{2017MNRAS.464.1211H} & 11 & TeV J2032+4130 & 47.9 & 21 \\
J2229+6114 &  51.6 & 77.9 & 2231 & 2 & Boomerang & 47.2 & 33 \\

\hline
\end{tabularx}

\label{tab:pulsar_properties}
\end{table}

\section{Acknowledgements}
This research is supported by grants from the U.S. Department of Energy Office of Science, the U.S. National Science Foundation and the Smithsonian Institution, and by NSERC in Canada. This research used resources provided by the Open Science Grid, which is supported by the National Science Foundation and the U.S. Department of Energy's Office of Science, and resources of the National Energy Research Scientific Computing Center (NERSC), a U.S. Department of Energy Office of Science User Facility operated under Contract No. DE-AC02-05CH11231. We acknowledge the excellent work of the technical support staff at the Fred Lawrence Whipple Observatory and at the collaborating institutions in the construction and operation of the instrument.

%This research is supported by grants from the U.S. Department of Energy Office of Science, the U.S. National Science Foundation and the Smithsonian Institution, and by NSERC in Canada. We acknowledge the excellent work of the technical support staff at the Fred Lawrence Whipple Observatory and at the collaborating institutions in the construction and operation of the instrument.

\bibliographystyle{JHEP}
\bibliography{./references}

%\begin{thebibliography}{99}
%\bibitem{myref}%3.1A, \textit{%T}, %J, %V, %p, %Y. %X\n 
%\end{thebibliography}

\end{document}